\documentstyle[12pt]{article}
\newcommand{\eq}[1]{eq.(\ref{#1})\ }
\newcommand{\ben}{\begin{equation}}
\newcommand{\een}{\end{equation}}
\newcommand{\bea}{\begin{eqnarray}}
\newcommand{\eea}{\end{eqnarray}}
\newcommand{\bear}{\begin{array}}
\newcommand{\enar}{\end{array}}
\newcommand{\bdm}{\begin{displaymath}}
\newcommand{\edm}{\end{displaymath}}
\newcommand{\nn}{\nonumber \\ }
\newcommand{\binomial}[2]{\left (\begin{array}{c} {#1}\\ {#2} \end{array}
\right )}

\newcommand{\pa}{\partial}

\newcommand{\N}{{\cal N}}

\newcommand{\g}{\gamma}
\newcommand{\G}{\Gamma}

\newcommand{\C}{\mbox{\hspace{1.24mm}\rule{0.2mm}{2.5mm}\hspace{-2.7mm} C}}

\newcommand{\Z}{\mbox{$Z\hspace{-2mm}Z$}}

\newcommand{\br}{\langle}
\newcommand{\kt}{\rangle}
\newcommand{\bra}[1]{\langle {#1}|}
\newcommand{\ket}[1]{|{#1}\rangle}

\newcommand{\dtp}[1]{\frac{d{#1}}{2\pi i}}

\newcommand{\vs}{\vspace}

\newcommand{\NP}[1]{Nucl.\ Phys.\ {\bf #1}}

\newcommand{\CMP}[1]{Commun.\ Math.\ Phys.\ {\bf #1}}

\newcommand{\MPL}[1]{Mod.\ Phys.\ Lett.\ {\bf #1}}

\newcommand{\IM}[1]{Invent.\ Math.\ {\bf #1}}
\newcommand{\SJNP}[1]{Sov. J. Nucl. Phys.\ {\bf #1}}

\begin{document}

\topmargin -5mm
\oddsidemargin 5mm

\begin{titlepage}
\setcounter{page}{0}
\begin{flushright}
NBI-HE-95-16\\
hep-th/9504127\\
April 1995\\
\end{flushright}

\vs{8mm}
\begin{center}
{\Large CONFORMAL BLOCKS FOR ADMISSIBLE}\\[.2cm]
{\Large REPRESENTATIONS IN $SL(2)$ CURRENT ALGEBRA}

\vs{8mm}
{\large Jens Lyng Petersen}\footnote{e-mail address:
 jenslyng@nbivax.nbi.dk},
{\large J{\o}rgen Rasmussen}
\footnote{e-mail address: jrasmussen@nbivax.nbi.dk}
{\large and Ming Yu}\footnote{e-mail address:
 yuming@nbivax.nbi.dk. Address after 1st October 1995:
Inst. of Theor. Phys., Academia Sinica, Beijing,
Peoples Republic of China}
\\[.2cm]
{\em The Niels Bohr Institute, Blegdamsvej 17, DK-2100 Copenhagen \O,
Denmark}\\[.5cm]

\end{center}

\vs{8mm}
\centerline{{\bf{Abstract}}}

Despite considerable work in the literature on $N$-point correlators in
2-d conformal WZNW models based on affine $\widehat{SL}(2)_k$, either by
using the Wakimoto construction or by directly solving the
Knizhnik-Zamolodchikov equations, most published results pertain to integrable
representations with $t=k+2$ integer and all primary fields having integer or
half integer spin. Results for admissible representations corresponding to
$t=k+2=p/q$ rational, appear to be
rather incomplete,
despite their potential interest in various connections, notably in connections
to non-critical string theory via
$2-d$ gravity based on hamiltonian reductions. Indeed, surprisingly, even
the fusion rules remain a subject of discussion. The reason for this state of
affairs may be traced to the need in the free field Wakimoto construction
for introducing a second screening charge as discussed by Bershadsky and
Ooguri,
one which depends on fractional powers of free fields, and such entities have
until now eluded
a consistent interpretation in terms of Wick contractions. In this
paper we develop the techniques necessary to deal with these complications,
and we provide
explicit general integral representations for conformal blocks on the sphere.
They turn out to have the structure expected from the operator formalism and
one, which renders their consistency check straight forward.
We further discuss fusion rules, and as a check we verify explicitly that our
conformal blocks satisfy the Knizhnik-Zamolodchikov equations and are
projectively invariant.
\end{titlepage}
\newpage
\renewcommand{\thefootnote}{\arabic{footnote}}
\setcounter{footnote}{0}
\section{Introduction}
$N$-point correlators of 2-$d$ conformal WZNW theories based on affine
$\widehat{SL}(2)_k$, with $k=$ level, have been much studied
already. They are typically constructed, either by applying the free field
realization of Wakimoto \cite{Wak}, from which results have been given for
example in refs. \cite{BF,ATY,FGPP,D90}, or by solving the
Knizhnik-Zamolodchikov
equations \cite{KZ}, from which results have been given for example in refs.
\cite{KZ,FZ,A,CF,SV}. Recently the structure of solutions of the
Knizhnik-Zamolodchikov equations on higher genus Riemann surfaces has been
reexamined \cite{FV}. The results given in these various pieces of works are
quite complete as far as unitary, integrable representations \cite{GW}
are concerned, but appear surprisingly incomplete for the general case,
including admissible representations.
Partial and conjectural results not
based on the free field realization were given in \cite{FGPP}. In particular
these authors made the interesting conjecture, that minimal model conformal
blocks are obtained from the affine $SL(2)$ blocks by a simple substitution,
one
of identifying the $x_i$-variables related to the $i$'th $SL(2)$
representation (and to be
introduced below), with the Koba-Nielsen variables, $z_i$.
Based on this conjecture an attempt was made to solve the
Knizhnik-Zamolodchikov
equations in a power series of $(x_i-z_i)$. In contrast we shall find complete
integral expressions based on the free field realization and providing
exact solutions to the KZ equations.
Using that realization, the conjecture may now be addressed. In this
paper we have merely checked it's validity in a few examples. We intend to
come back to a detailed discussion of this and related issues of hamiltonian
reduction elsewhere.

In general the WZNW theory is characterized by the level, $k$, or equivalently
by $t=k+2$ (for $\widehat{SL}(2)_k$). Then  degenerate primary fields
exist for representations characterized by spins, $j_{rs}$, given by
\cite{KK,MFF}
\ben
2j_{rs}+1=r-st
\een
with $r,s$ integers. However, previous results can be characterized as
pertaining only to the special case, $s=0$, which is the full case only for
integrable representations.
The reason for this restriction is fairly natural, since (see sect. 2)
the screening charge usually employed in the free field realization is capable
of screening just such primary fields. In fact, a possible second screening
operator, capable of screening the general case was proposed by Bershadsky
and Ooguri \cite{BO}, but since it involved fractional powers of the free
ghost fields, it apparently has remained unknown how to make use of that
screening operator.

In the present paper we overcome this difficulty by showing how the techniques
of fractional calculus \cite{MR} (briefly described in an appendix)
naturally provides a solution. As a result we are able to
present general integral formulas for the $N$-point conformal blocks on a
sphere, and
by using standard sewing techniques it should be possible to generalize those
to an arbitrary Riemann surface. Our own motivation lies in our wish to
make use of these results to cover the case of admissible representations
\cite{KW} corresponding to rational values of $t$, since these are the ones
relevant for treating conformal minimal matter using hamiltonian reduction
\cite{BO} coupled to 2-$d$ gravity along the lines of \cite{HY,AGSY}. Once
this technique is worked out it would be interesting to generalize to higher
groups and supergroups in order to be able to treat more general non-critical
string theory. The technique we present here appears directly amenable of such
generalizations.

In sect. 2 we define our notation and introduce the relation to
fractional calculus. In sect. 3
we show how to obtain the three point function and we derive fusion rules
for admissible representations and
compare with results already in the literature. In sect. 4 we derive the
$N$-point function and make comments on comparison with known results.
In sect. 5 we provide examples concerning 4-point functions.
In sect.
6 we prove that the $N$-point functions satisfy the
Knizhnik-Zamolodchikov equations. It appears from comparing with known
solutions
in the mathematics literature \cite{SV},
that our formulas involving auxiliary integrations represent fairly
powerful ways of dealing with such solutions.
In fact, the structure of our blocks corresponds
very closely to the operator formalism, and any results worked out in the
latter can be verified in the former.
In sect. 7 we discuss the slightly non trivial way in which projective and
global $sl_2$ invariance of the correlators is established.
In sect. 8 we give a summary
and an outlook. In two appendices we briefly describe fractional calculus
\cite{MR}, and
we prove some non trivial consistency conditions for
the rules for Wick contractions we have derived based on fractional calculus.

\section{Notation for free field realization. Relation to fractional calculus}

The Wakimoto realization \cite{Wak} is based on the free scalar field,
$\varphi(z)$, and bosonic ghost fields, $(\beta(z),\gamma(z))$,
of dimensions $(1,0)$ which we take to have the following contractions
\ben
\varphi(z)\varphi(w)=\log(z-w), \ \ \ \beta(z)\gamma(w)=\frac{1}{z-w}
\een
We only consider one chirality of the fields. The $\widehat{SL}(2)_k$ affine
currents may then be represented as
\bea
J^+(z)&=&\beta(z)\nn
J^3(z)&=&-:\gamma\beta:(z)-\sqrt{t/2}\pa\varphi(z)\nn
J^-(z)&=&-:\gamma^2\beta:(z)+k\pa\gamma(z)-\sqrt{2t}\gamma\pa\varphi(z)\nn
t&\equiv&k+2\neq 0
\label{wakimoto}
\eea
They satisfy
\bea
J^+(z)J^-(w)&=&\frac{2}{z-w}J^3(w)+\frac{k}{(z-w)^2}\nn
J^3(z)J^\pm(w)&=&\pm\frac{1}{z-w}J^\pm(w)\nn
J^3(z)J^3(w)&=&\frac{k/2}{(z-w)^2}
\eea
The Sugawara energy momentum tensor is obtained as
\ben
T(z)=:\beta\pa\gamma:(z)+\frac{1}{2}:\pa\varphi\pa\varphi:(z)+
\frac{1}{\sqrt{2t}}\pa^2\varphi(z)
\een
with central charge
$$c=\frac{3k}{k+2}$$
Since we shall be dealing with nonunitary representations, $j$, of $SL(2)$
for which the weight, $m$, may assume infinitely many (integrally spaced)
values, it is most convenient to collect the multiplets of primary fields as
\cite{FZ}
\ben
\phi_j(z,x)=\sum_m\phi^m_j(z)x^{j-m}
\een
Such a primary field satisfies the following OPE's
\ben
J^a(z)\phi_j(w,x)=\frac{1}{z-w}J_0^a(w)\phi_j(w,x)
\een
where the $SL(2)$ representation is provided by the differential operators
\bea
J_0^a(z)\phi_j(z,x)&=&[J_0^a, \phi_j(z,x)]=D_x^a \phi_j(z,x)\nn
D_x^+&=&-x^2\pa_x+2xj\nn
D_x^3&=&-x\pa_x+j\nn
D_x^-&=&\pa_x
\eea
The primary field $\phi_j(w,x)$ defined in this way transforms covariantly
under both conformal transformations and loop projective transformations,
namely as an $h$ tensor field for the former, and a $-j$ tensor field for
the latter,
\bea
z&\rightarrow& f(z)\nn
x&\rightarrow& \frac{a(z)x+b(z)}{c(z)x+d(z)}
\eea
with $a(z)d(z)-b(z)c(z)=1$.
One easily verifies that the free field realization of the primary field may
be taken as \cite{FGPP}
\ben
\phi_j(z,x)=(1+\gamma(z)x)^{2j}:e^{-j\sqrt{2/t}\varphi(z)}:
\label{pfd}
\een
where, in general one should asymptotically expand $(1+\gamma(z)x)^{2j}$ as
\ben
(1+\gamma(z)x)_{(\alpha)}^{2j}=
\sum_{n\in \Z}\binomial{2j}{n+\alpha}(\gamma(z)x)^{n+\alpha}
\een
Here, the choice of the parameter $\alpha$ depends on the monodromy conditions
of the primary field  $\phi_j(z,x)$ around contours in $x$-space, and those in
turn depend on the other fields present in the correlator.

Let us add various remarks of a general nature. For further details on
fractional calculus we refer to Appendix A. The most important rule for
us will simply be
\ben
\pa^a_x x^b=\frac{\Gamma(b+1)}{\Gamma(b-a+1)}x^{b-a}
\een
Next, consider the fractional derivative of the exponential function:
\ben
D^a\exp (x)=\sum_{n\in\Z}\frac{1}{\Gamma(n-a+1)}x^{n-a}, \ \ a\in\C
\een
This represents a peculiar realization of the exponential function itself,
which
converges asymptotically for $|x|\rightarrow \infty$.
Formally the right hand side is invariant under
further differentiation corresponding to the fact that it
represents the original exponential
function. The representation may be better understood
by writing
\ben
\exp\{x\}=x^{-a}{[}\frac{e^x}{x^{-a}}{]}
\een
and then introducing for the last bracket a Fourier expansion with integer
powers of $x$ on a circle in the complex $x$-plane. On the circle
$${[}e^xx^a{]}$$
has a discontinuity which we may take to be for negative $x$.
The Fourier
expansion converges for large $|x|$ where the discontinuity becomes vanishingly
small. Thus we take
\ben
D^a\exp(x)=\exp(x)
\een
for any $a$. However, we shall find it convenient to use the fractional
derivative to represent a generating functional for the integrals
\ben
\lim_{R\rightarrow\infty}\oint_{RS^1}\dtp{u}\frac{e^u}{u^{a+1+n}}=
\frac{1}{\Gamma(a+1+n)}
\een
Different $a$'s give rise to representations or expansions of the
exponential function in which individual terms have different non-trivial
monodromies.

Similarly, since we shall need contractions with the operator
$$(1+\gamma(z)x)^{2j}$$
we shall find it convenient to represent the expansion of the associated
analytic function
$$f(z)\equiv (1+z)^{2j}$$
in several different ways. Indeed, from
$$(1+z)^{2j}=z^a{[}(1+z)^{2j}z^{-a}{]}$$
we may imagine that the last bracket is expanded in integer powers of $z$
in a way convergent on the unit circle (with suitable conditions on $a$) since
in fact this time the discontinuity is vanishing. This
renders many equivalent representation for the function:
\ben
(1+z)^{2j}=\sum_{n\in\Z}\binomial{2j}{n-a}z^{n-a}
\een
which are all equivalent in the sense of analytic function theory but which
correspond to expansions with different monodromies for the individual
terms.

When deciding on what expansion to adopt for the operator
$$(1+\gamma(z)x)^{2j}$$
we use the criterion, that {\em after} all Wick contractions have been
performed, powers of $\beta$ and $\gamma$ inside normal ordering signs are
non-negative {\em integers,} such as was illustrated above.
Only then are these terms having an obvious interpretation
when sandwiched between states. In other terms, the existence of external
states and other primary fields in the correlator decides what monodromies to
choose for individual terms in expansions. All of that will be illustrated
further below.

To calculate operator product expansion (OPE) of expressions involving
the $\beta(z)$ and $\gamma(w)$
fields, when either $\beta$ or $\gamma$ fields appear with integral powers,
it is clear that the following rules apply:
\bea
\beta(z)^n F(\gamma(w))&=&:(\beta(z)+\frac{1}{z-w}\pa_{\gamma(w)})^n
F(\gamma(w)
):\nn
\gamma(z)^n F(\beta(w))&=&:(\gamma(z)-\frac{1}{z-w}\pa_{\beta(w)})^n F(\beta(w)
):
\label{bgcon}
\eea
The two relevant screening charge currents are \cite{BO}
\bea
S_1(z)&=&\beta(z)e^{+\sqrt{2/t}\varphi(z)}\nn
S_2(z)&=&\beta(z)^{-t}e^{-\sqrt{2t}\varphi(z)}
\label{screen}
\eea
{}From now on we mostly leave out normal ordering signs around exponentials.
The screening currents are easily seen to have OPE's with the affine currents
that are either non-singular or form total derivatives (in particular this
will follow from our definitions of Wick contractions for fractional powers
of free fields below).

As we have seen, the $\gamma(z)$ field can be fractionally powered
in
asymptotic expansions of the primary field $\phi_j(z,x)$. Thus we need
insertions of the second screening current $S_2(z)$ in the correlators to
neutralize those. In such cases, the $\beta$ field has nontrivial monodromy
with respect to the primary field  $\phi_j(z,x)$.
Hence our proposal to deal with the awkward looking field,
$$\beta(z)^{-t}$$
consists in generalizing \eq{bgcon} as
\ben
G(\beta(z)) F(\gamma(w))=:G(\beta(z)+\frac{1}{z-w}\pa_{\gamma(w)})F(\gamma(w)):
\label{funcbgcon}
\een
where, the asymptotic expansions for $G(\beta(z)+\frac{1}{z-w}\pa_{\gamma(w)})$
and $F(\gamma(w))$ would depend on their monodromy conditions in the
$z$ and $w$
variables respectively. When applying \eq{funcbgcon} to our particular case,
where $F(\gamma)=(1+x\gamma)^{2j}$, we have
\bea
\beta(z)^{-t} F(\gamma(w))&=&:(\beta(z)+\frac{1}{z-w}\pa_{\gamma(w)})^{-t}
F(\gamma(w)):\nn
&=&\sum_{n=0}^\infty\binomial{-t}{n}:\beta^n(z)(z-w)^{t+n}\pa_{\gamma(w)}^{-t-n}
F(\gamma(w)):
\label{frbgcon}
\eea
and we see the need for fractional calculus \cite{MR}.

As an example of how the technique works we provide in the appendix an explicit
non-trivial proof that
\ben
(\beta^a(z)\gamma^a(w))(\beta^b(z)\gamma^b(w))=\beta^{a+b}(z)\gamma^{a+b}(w)
\een
Additionally, our explicit verification that our results for the $N$-point
functions satisfy the Knizhnik-Zamolodchikov equations may be viewed as a
check that fractional calculus
does indeed provide us with the requisite properties for Wick contractions as
defined in \eq{frbgcon}.

\section{The three point function and fusion rules}
Before considering correlators, we want to define our notations and
choices
as far as dual states are concerned \cite{FMS,D90}. We use the following mode
expansions and definitions:
\bea
\beta(z)=\sum_{n\in\Z}\beta_nz^{-n-1}&,&\gamma(z)=\sum_{n\in\Z}\gamma_nz^{-n}\nn
:\beta_n\gamma_m:&=&\left\{\begin{array}{lcl}
\beta_n\gamma_m&,&n<0\\
\gamma_m\beta_n&,&n\geq 0\end{array}\right.\nn
j(z)&=&-:\gamma(z)\beta(z): =+\pa\phi(z)\nn
\phi(z)\phi(z')&=&-\log(z-z')\nn
\varphi(z)\varphi(z')&=&+\log(z-z')\nn
\phi(z)&=&q_\phi+a_\phi\log z+\sum_{n\neq 0}\frac{j_n}{-n}z^{-n}\nn
\varphi(z)&=&q_\varphi +a_\varphi\log z +\sum_{n\neq 0}\frac{a_n}{-n}z^{-n}\nn
j_0&\equiv&a_\phi\nn
{[}a_\phi,q_\phi{]}&=&-1\nn
{[}a_\varphi,q_\varphi{]}&=&+1\nn
j(z)&=&\sum_{n\in\Z}j_nz^{-n-1}\nn
j_n^\dagger=-j_{-n}&,&a_n^\dagger=a_{-n}\nn
\gamma_n^\dagger=\gamma_{-n}&,&\beta_n^\dagger= -\beta_{-n}
\eea
The ket-vacuum, invariant under both projective $sl_2$ and affine zero-mode
$sl_2$, is $\ket{0}$ satisfying
\bea
\beta_n\ket{0}&=&0=\gamma_n\ket{0}=a_n\ket{0}=j_n\ket{0}, \ n>0\nn
\beta_0\ket{0}&=&0=a_\varphi\ket{0}=a_\phi\ket{0}\nn
\gamma_0\ket{0}&\neq& 0\nn
L_n\ket{0}&=&0, \ \ n\geq -1
\eea
with
\ben
L_n=\sum_{m\in\Z}(-m:\beta_{n-m}\gamma_m:+
\frac{1}{2}:a_{n-m}a_{m}:)-(n+1)\sqrt{\frac{1}{2t}}a_n
\een
Correspondingly the $sl_2$ invariant bra-vacuum, $\bra{sl_2}$, satisfies
\bea
\bra{sl_2}L_n&=&0, \ n\leq 1\nn
\bra{sl_2}\beta_n&=&0,\ n\leq 0\nn
\bra{sl_2}\gamma_n&=&0, \ n\leq -1\nn
\bra{sl_2}j_0&=&\bra{sl_2}a_\phi =\bra{sl_2}\nn
\bra{sl_2}a_\varphi&=&\sqrt{\frac{2}{t}}\bra{sl_2}\nn
\bra{sl_2}0\kt&=&0
\eea
The last equality is due to the fact that the bra-vacuum defined above carries
different charges comparing to the ket-vacuum. In what follows we shall
define another bra-vacuum with all the charges at infinity screened. This
will be the dual vacuum we are mostly going to use in calculating the
correlators.

We define the {\em dual vacuum state} in the WZNW free field
realization, $\bra{0}$, as
\ben
\bra{0}=\bra{sl_2}e^{-q_\phi}e^{\sqrt{2/t}q_\varphi}
\een
It satisfies:
\bea
\bra{0}0\kt&=&1\nn
\bra{0}\gamma_0&=&0\nn
\bra{0}\beta_0&\neq&0\nn
\bra{0}a_\varphi&=&0\nn
\bra{0}j_0=\bra{0}a_\phi&=&0\nn
\bra{0}\beta(z)\gamma(z')\ket{0}&=&\frac{1}{z-z'}
\eea

{}From the dual vacuum we construct dual bra-states of lowest $SL(2)$ weight
\ben
\bra{j}=\bra{0}e^{j\sqrt{2/t}q_\varphi}
\een
This state indeed satisfies the conditions for being a {\em lowest weight
state}
of the affine algebra:
\bea
\bra{j}J^3_0&=&j\bra{j}\nn
\bra{j}J^-_n&=&0, \ n\leq 0\nn
\bra{j}J^3_n&=&0, \ n<0\nn
\bra{j}J^+_n&=&0, \ n<0
\eea
For the corresponding ket states
\ben
\ket{j}=e^{-j\sqrt{2/t}q_\varphi}\ket{0}
\een
we have
\ben
\bra{j}j\kt=1
\een
and this ket state is similarly a {\em highest weight state}
of the affine algebra.
We notice that
\ben
\bra{0}J^+_0=\bra{0}\beta_0\neq 0
\een
Thus we are performing all calculations with an $sl_2$ non-invariant
bra-vacuum.
This gives rise to some complications when we wish to prove projective and
global $sl_2$ invariance of our correlators. We shall explicitly demonstrate
in sect. 7 that the above state, $\bra{0}\beta_0$, while not being zero
is in fact a BRST exact state in the sense of Felder \cite{F},
and that therefore it
must be expected to decouple from all correlators. This decoupling we then
verify.

Let us now consider the evaluation of the (chiral) three point function
\ben
\bra{j_3}\phi_{j_2}(z,x)\ket{j_1}
\een
where, the dual bra, $\bra{j_3}$, and the ket, $\ket{j_1}$, are defined
above. Using the
free field realizations of $\phi_{j_2}(z,x)$, \eq{pfd}, the three point
function may be evaluated only provided the ``momenta" may be
screened away in the standard way \cite{DF}, and correspondingly
$\phi_{j_2}(z,x)$ is replaced by the intertwining field,
$(\phi_{j_2}(z,x))_{j_1}^{j_3}$,
which maps a $j_1$ highest weight module into a $j_3$ highest weight module.
Following Felder \cite{F,BF}, but using the two
screening charges of \cite{BO} in \eq{screen} instead, we are led to consider
the intertwining field
\bea
(\phi_{j_2}(z,x))_{j_1}^{j_3}&=&\oint\prod_{j=1}^s
\dtp{v_j}\prod_{i=1}^r\dtp{u_i}\phi_{j_2}(z,x)P(u_1,...,u_r;v_1,...,v_s)\nn
P(u_1,...,u_r;v_1,...,v_s)&=&\prod_{j=1}^s\beta^{-t}(v_j)
e^{-\sqrt{2t}\varphi(v_j)}\prod_{i=1}^r\beta(u_i)e^{\sqrt{2/t}\varphi(u_i)}
\label{intertw}
\eea
This requires that
\ben
j_1+j_2-j_3=r-st
\label{3rs}
\een
with $r$ and $s$ non negative integers.
It is trivial using well known techniques to perform the $\varphi$ part of the
Wick contractions. Hence we concentrate on explaining how to perform the
$\beta\gamma$ part.
First we have to determine the asymptotic expansion in $\gamma$ within
$\phi_{j_2}(z,x)$. By projective invariance of the three point function,
$x$ could be fractionally powered, when $s$ in \eq{3rs} is nonzero:
\ben
\bra{j_3}(\phi_{j_2}(z,x))_{j_1}^{j_3}\ket{j_1}\propto x^{r-st}
\een
Hence we should expand asymptotically,
\bea
\phi_{j_2}(z,x)&=&(1+\gamma(z)x)_{(-st)}^{2j_2}e^{-j_2\sqrt{2/t}\varphi(z)}\nn
               &=&\sum_{n\in \Z}\binomial{2j_2}{n-st}(\gamma(z)x)^{n-st}
e^{-j_2\sqrt{2/t}\varphi(z)}
\eea
and
\bea
&&\beta(w)^{-t}(1+\gamma(z)x)_{(-st)}^{2j_2}\nn
&=&:(\beta(w)+\frac{1}{w-z}\pa_{\gamma(z)})^{-t}
\sum_{m\in \Z}\binomial{2j_2}{m-st}(\gamma(z)x)^{m-st}:\nn
&=&\sum_{n=0}^\infty\binomial{-t}{n}:\beta^n(w)(w-z)^{t+n}\pa_{\gamma(z)}^{-t-n}
\sum_{m\in \Z}\binomial{2j_2}{m-st}(\gamma(z)x)^{m-st}:\nn
&=& \sum_{n=0}^\infty\binomial{-t}{n}:\beta^n(w)(w-z)^{t+n}
\frac{\Gamma(2j_2+1)}{\Gamma(2j_2+t+n+1)}
(1+\gamma(z)x)_{((1-s)t)}^{2j_2+t+n}:x^{-t-n}
\label{preexp}
\eea
Similarly, we deduce
\bea
&&\prod_{i=1}^{s}\beta(w_i)^{-t}(1+\gamma(z)x)_{(-st)}^{2j_2}\nn
&=&:\prod_{i=1}^{s}(\beta(w_i)+\frac{1}{w_i-z}\pa_{\gamma(z)})^{-t}
\sum_{m\in \Z}\binomial{2j_2}{m-st}(\gamma(z)x)^{m-st}:\nn
&=&:\prod_{i=1}^{s}\sum_{n_i=0}^\infty
\binomial{-t}{n_i}\beta^{n_i}(w_i)(w_i-z)^{t+n_i}\pa_{\gamma(z)}^{-t-n_i}
\sum_{m\in \Z}\binomial{2j_2}{m-st}(\gamma(z)x)^{m-st}:\nn
&=&:\prod_{i=1}^{s}\sum_{n_i=0}^\infty
\binomial{-t}{n_i}\beta^{n_i}(w_i)(w_i-z)^{t+n_i}
\frac{\Gamma(2j_2+1)}{\Gamma(2j_2+st+\sum_i n_i+1)}\nn
&\cdot&
(1+\gamma(z)x)^{2j_2+st+\sum_i n_i}:x^{-st-\sum_i n_i}
\label{preexp1}
\eea
Notice that in these equations $\beta$ and $\gamma$ appear within normal
ordering signs with integral powers.

Eqs.(\ref{preexp}) and (\ref{preexp1}) suggest that we consider some kind of
generating function, which looks like the exponential function,
\ben
F(u)=\sum_{n\in \Z}\frac{1}{\Gamma(n-a+1)}
(1+\gamma(z)x)_{(\alpha)}^{n-a}u^{-n+a}
\een

We shall find it useful to use the following rather trivial
identity:
\ben
(1+\gamma(z)x)^{2j}=\Gamma(2j+1)\oint_0\dtp{u}\frac{1}{u}(u^{-1}D)^{-2j}
\exp{[}(1+\gamma(z)x)/u{]}
\een
where $D$ converts the exponential function into the derivative of that
function, in particular it acts on and only on
the entire argument of that function. We now prove the following

{\bf Lemma}
\ben
\beta^a(w)\exp{[}(1+\gamma(z)x)/u{]}=:(\beta(w)+\frac{x/u}{w-z})^aD^a
\exp{[}(1+\gamma(z)x)/u{]}:
\label{lemma}
\een

{\em Proof of lemma:}
\bea
&&\beta^a(w)\exp{[}(1+\gamma(z)x)/u{]}\nn
&=&\sum_{k,m,n\in\Z}\binomial{a}{m}:\beta^m(w)(w-z)^{m-a}\pa_{\gamma(z)}^{a-m}
\frac{1}{\Gamma(n+1)}\binomial{n}{k}(\gamma(z)x)^{n-k}:u^{-n}\nn
&=&\sum_{k,m,n\in\Z}\binomial{a}{m}:\beta^m(w)(w-z)^{m-a}\gamma^{n-k-a+m}(z):\nn
&\cdot&\frac{1}{\Gamma(k+1)}\frac{1}{\Gamma(n-k-a+m+1)}x^{n-k-a+m}x^{a-m}u^{-n}
\nn
&=&\sum_{k,m,N\in\Z}\binomial{a}{m}:\beta^m(w)(w-z)^{m-a}x^{a-m}u^{m-a}
\gamma^{-a+N-k}(z):\nn
&\cdot&\binomial{-a+N}{k}\frac{1}{\Gamma(-a+N+1)}x^{-a+N-k}u^{a-N}\nn
&=&:(\beta(w)+\frac{x/u}{w-z})^aD^a\exp{[}(1+\gamma(z)x/u){]}:
\eea
Q.E.D.

We may now calculate the $\beta\gamma$ parts of the contractions in the
3-point function:
\bea
&&\prod_{i=1}^r\beta(u_i)\prod_{j=1}^s\beta^{-t}(v_j)
(1+\gamma(z)x)^{2j}\nn
&=&\oint_0\dtp{u}\frac{u^{2j}}{u}
:\prod_{i=1}^r{[}\beta(u_i)+\frac{x/u}{u_i-z}{]}
\prod_{j=1}^s{[}\beta(v_j)+\frac{x/u}{v_j-z}{]}^{-t}\nn
&\cdot&D^{-2j+r-st}\exp\{\frac{1+\gamma(z)x}{u}\}:
\Gamma(2j+1)
\eea
When inserted between $\bra{j_3}$ and $\ket{j_1}$ to produce
$W_3^{\beta\gamma}$, the $\beta\gamma$ part of the three point function, we
effectively put $\beta\equiv 0\equiv \gamma$ whereupon the $u$-integration
becomes trivial, and we find the result
\ben
W_3^{\beta\gamma}=\frac{\Gamma(2j_2+1)}{\Gamma(2j_2-r+st+1)}x^{r-st}
\oint\prod_{i=1}^r\dtp{u_i}(u_i-z)^{-1}\prod_{j=1}^s\dtp{v_j}(v_j-z)^t
\een
Here we mean to employ the contours discussed by Felder \cite{F}. The $\varphi$
part of the 3-point function is standard.
We may put $z=1=x$ using global projective and global $SL(2)$
invariance. We only give the result for admissible representations:
\bea
t&=&p/q\nn
2j_i+1&=&r_i-s_it\nn
1&\leq& r_i\leq p-1 \nn
0&\leq& s_i\leq q-1
\eea
where $(p,q)=1$, and $p,q\in\N$.
\bea
W_3&=&\frac{\Gamma(2j_2+1)}{\Gamma(2j_2-r+st+1)}\nn
&\cdot&\oint\prod_{i=1}^r\dtp{u_i}\prod_{j=1}^s\dtp{v_j}\prod_{i_1<i_2}
(u_{i_1}-u_{i_2})^{2/t}\prod_{j_1<j_2}(v_{j_1}-v_{j_2})^{2t}
\prod_{i,j}(u_i-v_j)^{-2}\nn
&\cdot&\prod_{i=1}^ru_i^{(1-r_1)/t+s_1}(1-u_i)^{(1-r_2)/t+s_2-1}
\prod_{j=1}^sv_j^{r_1-1-s_1t}(1-v_j)^{r_2-1-(s_2-1)t}
\eea
This integral is exactly of the Dotsenko-Fateev form \cite{DF} and, using the
Felder contours, may be written down as \cite{F}
\bea
W_3&=&\frac{\Gamma(2j_2+1)}{\Gamma(j_2+j_3-j_1+1)}
e^{i\pi r(r+1-2r_1)/t}e^{i\pi ts(s-1-2s_1)}t^{rs} \nn
&\cdot&\prod_{j=1}^r\frac{(1-e^{2\pi i(r_1-j)/t})(1-e^{2\pi i j/t})}
{1-e^{2\pi i/t}}
\prod_{j=1}^s\frac{(1-e^{2\pi it(s_1+1-j)})(1-e^{2\pi itj})}
{1-e^{2\pi it}}\nn
&\cdot&\prod_{i=1}^r\frac{\Gamma(i/t)}{\Gamma(1/t)}\prod_{i=1}^s
\frac{\Gamma(it-s)}{\Gamma (t)}\nn
&\cdot&\prod_{i=0}^{r-1}\frac{\Gamma(s_1+1+(1-r_1+i)/t)\Gamma(s_2+(1-r_2+i)/t)}
{\Gamma(s_1+s_2+1-2s+(r-r_1-r_2+i+1)/t)}\nn
&\cdot&\prod_{i=0}^{s-1}\frac{\Gamma(r_1-r+(i-s_1)t)\Gamma(r_2-r+(1-s_2+i)t)}
{\Gamma(r_1-r+r_2+(s-s_1-s_2+i)t)}
\eea
The analysis of this expression in terms of fusion rules is standard \cite{F}.
The result may be written as follows:
\bea
1+|r_1-r_2|\leq&r_3&\leq p-1-|r_1+r_2-p|\nn
|s_1-s_2|\leq&s_3&\leq q-1-|s_1+s_2-q+1|
\label{FI}
\eea
The first line of these fusion rules is
well known for the case, $q=1$, of integrable
representations, and it was obtained in the general case in \cite{BF}. The
second was obtained by Awata and Yamada \cite{AY}
by considering the conditions for decoupling of null-states,
and by Feigin and Malikov \cite{FM} by cohomological methods. In
addition these authors provide a fusion rule ((II) for \cite{AY}, (I) for
\cite{FM}), which we do not get in the
free field realization. We do not know if there exist conformal field
theories with non vanishing couplings respecting those.

We conclude this section by making a comment on the possible $SL(2)$
representations carried by the intertwining field
$(\phi_{j_2}(z,x))_{j_1}^{j_3}$. Clearly, on the ket-vacuum,
\bea
(\phi_{j_2}(z,x))_{j_1}^{j_3}\ket{0} &=&
(\phi_{j_2}(z,x))_{0}^{j_2}\ket{0}\delta_{j_1,0}\nn
&=&e^{xJ_0^-}e^{zL_{-1}}\ket{j_2}\delta_{j_1,0}
\eea
is in a highest weight representation of the $SL(2)$ current algebra with
the highest weight state $\ket{j_2}$. On the other hand, as it will be
explicitly checked in sect. 4, on the dual vacuum state,
\bea
\bra{0}(\phi_{j_2}(z,x))_{j_1}^{j_3}z^{\frac{2j_2(j_2+1)}{t}}x^{-2j_2} &=&
\bra{0}(\phi_{j_2}(z,x))_{j_2}^{0}z^{\frac{2j_2(j_2+1)}{t}}x^{-2j_2}\nn
&=&\bra{j_2}e^{L_{1}/z}e^{-J_0^+/x}
\eea
is in a lowest weight representation of the $SL(2)$ current algebra with
the lowest weight state $\bra{j_2}$. When sandwiched in the middle of the
correlator, however, the intertwining field $(\phi_{j_2}(z,x))_{j_1}^{j_3}$
could carry representations belonging to the continuous series
of the $SL(2)$ algebra, in
which neither a highest weight nor a lowest weight state exists, when both
$j_1+j_2-j_3$ and $j_1-j_2-j_3$ are non-integers. This means we decompose
$\phi_{j_2}(z,x)_{j_1}^{j_3}$ into $J^3_0$ eigenstates,
\bea
[J_0^3,(\phi_{j_2,\lambda}(z,x))_{j_1}^{j_3}]&=&\lambda
(\phi_{j_2,\lambda}(z,x))_{j_1}^{j_3}\nn
\lambda&\neq& \pm j_2 \ (\mbox{mod}) \ 1
\eea
Although in that case $(\phi_{j_2}(z,x))_{j_1}^{j_3}$ does not correspond to a
highest weight representation, it maps a $j_1$ highest weight
representation to a $j_3$ highest weight representation.

\section{The most general $N$-point function for degenerate representations
on the sphere}

We wish to evaluate the conformal block
\ben
W_N=\bra{j_N}{[}\phi_{j_{N-1}}(z_{N-1},x_{N-1}){]}^{j_N}_{\iota_{N-2}}
...{[}\phi_{j_{n}}(z_{n},x_n){]}^{\iota_{n}}_{\iota_{n-1}}...
{[}\phi_{j_{2}}(z_2,x_2){]}^{\iota_2}_{j_1}\ket{j_1}
\een
Thus we have primary fields (chiral vertex operators) at points
$$z_1=0,z_2,...,z_{N-1},z_N=\infty$$
having $x$ values
$$x_1=0,x_2,...,x_{N-1},x_N=\infty$$
We parametrize
\bea
j_1+j_2-\iota_2&=&\rho_2-\sigma_2 t\nn
\iota_2+j_3-\iota_3&=&\rho_3-\sigma_3 t\nn
&\vdots&\nn
\iota_{n-1}+j_n-\iota_n&=&\rho_n-\sigma_n t\nn
&\vdots&\nn
\iota_{N-2}+j_{N-1}-j_N&=&\rho_{N-1}-\sigma_{N-1}t\nn
2j_i+1&=&r_i-s_it
\eea
with $\sigma_n,\rho_n$ non negative integers.
We then get for the $\beta\gamma$ part of the correlator, denoting by
$w(n,i)$ and $v(n,k)$ the positions of the $i$'th and the $k$'th screening
currents of the first and second kinds respectively in \eq{screen} around
the $n$'th primary field:
\bea
W^{\beta\gamma}_N&=&\bra{j_N}(1+x_n\gamma(z_n))^{2j_n}
\prod_{k=1}^{\sigma_n}{[}\beta(v(n,k)){]}^{-t}
\prod_{n=2}^{N-1}\prod_{i=1}^{\rho_n}\beta(w(n,i))
\ket{j_1}\nn
&=&\oint_0 \prod_{n=2}^{N-1}\dtp{u_n}\prod_{i=1}^{\rho_n}
\prod_{k=1}^{\sigma_n}\nn
&&\bra{j_N}:\left ( \beta(w(n,i))+\frac{x_n/u_n}{w(n,i)-z_n}\right )
\left(\beta(v(n,k))+\frac{x_n/u_n}{v(n,k)-z_n}\right )^{-t}\nn
&\cdot& u_n^{2j_n}D_n^{-2j_n+\rho_n-\sigma_nt}
\exp\{\frac{1+x_n\gamma(z_n)}{u_n}\}\Gamma(2j_n+1)\frac{1}{u_n}:\ket{j_1}\nn
&=&\oint_0\prod_{n=2}^{N-1}\dtp{u_n}\prod_{i=1}^{\rho_n}
\left(\sum_{\ell=2}^{N-1}
\frac{x_\ell/u_\ell}{w(n,i)-z_\ell}\right )\prod_{k=1}^{\sigma_n}
\left(\sum_{\ell=2}^{N-1}\frac{x_\ell/u_\ell}{v(n,k)-z_\ell}\right)^{-t}\nn
&\cdot&u_n^{2j_n-1}D_n^{-2j_n+\rho_n-\sigma_nt}\exp\{\frac{1}{u_n}\}
\Gamma(2j_n+1)
\eea
where we have used the techniques already developed for the three
point function.
In particular, in the second equality we applied the lemma of the previous
section, and in the last equality we kept doing that until normal ordering
signs surround all operators, at which point the calculation is completed by
putting $\beta$'s and $\gamma$'s under normal ordering signs equal to zero.
Conforming with the discussion in the previous section we may also throw
away all derivatives on the exponential (they are with respect to the full
argument of the exponential), but their presence serve to remind us in some
cases, what representations would conveniently be used. In the sequel we drop
these derivatives.
It is straightforward to write down the contribution from the $\varphi$ part
of the free field realization. It is
\bea
W^\varphi_N&=&\prod_{1\leq m<n\leq N-1}(z_m-z_n)^{2j_mj_n/t}
\prod_{n=2}^{N-1}\prod_{i=1}^{\rho_n}\prod_{m=1}^{N-1}(w(n,i)-z_m)^{-2j_m/t}\nn
&\cdot&\prod_{n=2}^{N-1}\prod_{k=1}^{\sigma_n}\prod_{m=1}^{N-1}
(v(n,k)-z_m)^{2j_m}\prod_{(n,i)<(n',i')}(w(n,i)-w(n',i'))^{2/t}\nn
&\cdot&\prod_{(n,k)<(n',k')}(v(n,k)-v(n',k'))^{2t}\prod_{(n,i),(n',k)}
(w(n,i)-v(n',k))^{-2}
\eea
Here we have introduced a rather arbitrary ordering of indices, for example
as
\ben
(n,i)<(n',i')
\een
if either $n<n'$ or $n=n', \ \ \ i<i'$.

Let us summarize our findings in a more compact notation: Let
\bea
M&=&\sum_{m=2}^{N-1}(\rho_m+\sigma_m)\nn
w_i&&i=1,..., M
\eea
collectively denote the position of all screening charges:
$$\{w_i\}=\{w(n,i), v(n,k)\}$$
Further, let
\bea
k_i&=&\left\{\begin{array}{rl}
-1&i=1,...,\sum_m\rho_m\\
t&i=\sum_m\rho_m+1,...,M
\end{array}\right. \nn
B(w_i)&\equiv&\sum_{\ell=1}^{N-1}\frac{x_{\ell}/u_{\ell}}{w_i-z_\ell}
\label{B}
\eea
(here $x_1=0$).
Then the {\em integrand} of the $N$-point function is given by (we use the same
letters for the integrated expressions, we hope this will not cause confusion)
\ben
W_N=W_BW^\varphi_NF
\label{npoint}
\een
with
\bea
W_N^{\beta\gamma}&=&W_BF\nn
W_B&=&\prod_{i=1}^M B(w_i)^{-k_i}\nn
W^\varphi_N&=&\prod_{m<n}(z_m-z_n)^{2j_mj_n/t}\prod_{i=1}^M\prod_{m=1}^{N-1}
(w_i-z_m)^{2k_ij_m/t}\prod_{i<j<M}(w_i-w_j)^{2k_ik_j/t}\nn
F&=&\prod_{m=2}^{N-1}\Gamma(2j_m+1)u_m^{2j_m-1}e^{\frac{1}{u_m}}
\eea
This integrand is to be integrated over the $u_m$'s along contours enclosing
$u_m=0$, and over the $w_i$'s along the Felder contours.

We believe the above general closed expression for integral representation
of the $N$-point function to be useful for further development, in particular
integrations over the auxiliary variables,
$u_\ell$, $\ell=2,...,N-1$ seem tractable as they stand.
If for some reason, one needs to get rid of these integrations, it is not
too difficult. As an example, we provide an explicit form for the result
for integrable representations. First we define the following notation:

Let
\bea
J_N&=&\{2,3,...,N-1\}\nn
I_N&=&\{(n,i)|n=2,...,N-1, \ i=1,2,...,\rho_n\}\nn
{\cal F}_N&=&\{\mbox{maps, $f$, from $I_N$ to $J_N$}\}
\eea

For $t=k+2$ integer, all $\sigma$'s are $=0$. In this case we may then write
\bea
W^{\beta\gamma}_N&=&\oint_0\prod_{n=2}^{N-1}\dtp{u_n}u_n^{2j_n}
\frac{e^{1/u_n}}{u_n}(2j_n)!\prod_{(n,i)\in I_N}\left(\sum_{\ell=2}^{N-1}
\frac{x_\ell/u_\ell}{w(n,i)-z_\ell}\right)\nn
&=&\oint_0\prod_{n=2}^{N-1}\dtp{u_n}u_n^{2j_n}\frac{e^{1/u_n}}{u_n}(2j_n)!
\sum_{f\in {\cal F}_N}\prod_{(n,i)\in I_N}
\left (\frac{x_{f(n,i)}/u_{f(n,i)}}{w(n,i)-z_{f(n,i)}}\right)\nn
&=&\oint_0\prod_{n=2}^{N-1}\dtp{u_n}u_n^{2j_n}\frac{e^{1/u_n}}{u_n}(2j_n)!\nn
&&\sum_{f\in {\cal F}_N}\prod_{\ell\in J_N}(x_\ell/u_\ell)^{|f^{-1}(\ell)|}
\prod_{(n,i)\in f^{-1}(\ell)} (w(n,i)-z_{\ell})^{-1}\nn
&=&\sum_{f\in {\cal F}_N}\prod_{\ell\in J_N}x_\ell^{|f^{-1}(\ell)|}
\frac{(2j_\ell)!}{(2j_\ell-|f^{-1}(\ell)|)!}
\prod_{(n,i)\in f^{-1}(\ell)} (w(n,i)-z_{\ell})^{-1}
\eea
A similar but even more complicated sum formula obtains in the general case.

We now have two ways of calculating the conformal block corresponding to
$N$ primaries. The way so far described is by using (part of) projective
invariance (to be discussed further later)
and global $sl_2$ invariance to work it out as
\bea
&&W^{(I)}_N(z_N=\infty,x_N=\infty,z_{N-1},x_{N-1},...,z_2,x_2,z_1=0,x_1=0) \nn
&=&\bra{j_N}{[}\phi_{j_{N-1}}(z_{N-1},x_{N-1}){]}^{j_N}_{\iota_{N-2}}...
{[}\phi_{j_2}(z_2,x_2){]}^{\iota_2}_{j_1}\ket{j_1}
\eea
However, obviously, we may also use our technique to evaluate the same
$N$-point conformal block as
\bea
&&W^{(II)}_N(z_N,x_N,z_{N-1},x_{N-1},...,z_2,x_2,z_1,x_1)\nn
&=&\bra{0}{[}\phi_{j_N}(z_N,x_N){]}^0_{j_N}
{[}\phi_{j_{N-1}}(z_{N-1},x_{N-1}){]}^{j_N}_{\iota_{N-2}}...\nn
&&...{[}\phi_{j_2}(z_2,x_2){]}^{\iota_2}_{j_1}
{[}\phi_{j_1}(z_1,x_1){]}^{j_1}_0 \ket{0}
\eea
We now want to demonstrate that up to normalization these expressions are
equivalent in the appropriate limits.
This procedure has also been mentioned
in section 3 when discussing different representations mapped into one another
by the intertwining field. Here is the direct check.
Notice that the second form involves more screening charges around the last
field than the first one. We shall see that these extra screenings give
rise to a constant contribution in the limit $z_N,x_N\rightarrow\infty$.
But for finite $z_N$ and $x_N$, unlike in the case of the conformal minimal
models, there does not seem to be any
simple way of getting a conjugate field, which would get
rid of the extra screening charges.

It is clear that in the limit $z_1,x_1\rightarrow 0$ the second formulation
coincides with the first. In
particular, the second formulation involves no extra screening operators.
Thus we shall concentrate on the limit $z_N,x_N\rightarrow\infty$.

As before, we let $w_i$ denote the position of screening operators {\em
in the first
case,} $W_N^{(I)}$, and we let $i$ run over the same set as in the first case.
Further,
we let $z_n,x_n$ denote the arguments as in the first case and $n=1,2,...,N-1$
runs over the same set as in the first case. The new feature in the second
case is:

(i) the appearance of
$$j_N+j_N-0=r_N-1-s_Nt=\rho_N-\sigma_Nt$$
extra screening operators, the positions of which we denote by
$$w^N_{i_N}, \ i_N=1,...,\rho_N+\sigma_N$$
$$k^N_{i_N}=\left\{\begin{array}{rcl}
-1&,&i_N=1,...,\rho_N\\
t&,&i_N=\rho_N+1,...,\rho_N+\sigma_N
\end{array}\right.$$
and

(ii) an extra $u$-integration over a variable we call $u_N$.

Then we want to consider the limit as $z_N,x_N\rightarrow\infty$ (letting
$W_N^{(I)}$ stand for the integrand in an appropriate way) of
\bea
&&z^{\frac{2j_N(j_N+1)}{t}}x^{-2j_N}W_N^{(II)}\nn
&=&z^{\frac{2j_N(j_N+1)}{t}}x^{-2j_N}\Gamma(2j_N+1)
\prod_{i_N}B^{-k^N_{i_N}}(w^N_{i_N})
\prod_{i_N<j_N}(w^N_{i_N}-w^N_{j_N})^{2k^N_{i_N}k^N_{j_N}/t}\nn
&\cdot&\prod_{i,i_N}(w^N_{i_N}-w_i)^{2k^N_{i_N}k_i/t}
\prod_{i_N}(z_N-w^N_{i_N})^{2k^N_{i_N}j_N/t}\prod_{i_N,n}(w^N_{i_N}-z_n)
^{2k^N_{i_N}j_n/t}\nn
&\cdot&\prod_n(z_N-z_n)^{2j_Nj_n/t}\prod_i(z_N-w_i)^{2j_Nk_i/t}
u_N^{2j_N-1}e^{1/u_N}\dtp{u_N}W_N^{(I)}\prod_{i_N}dw^N_{i_N}
\eea
where the function $B(w)$ is defined with one more term than for case (I), cf.
\eq{B}.
We now use that
\bea
-\sum_{i_N}k^N_{i_N}&=&\rho_N-\sigma_Nt= 2j_N\nn
\sum_ik_i&=&-\sum_nj_n+j_N
\eea
In the limit, $z_N\rightarrow\infty, x_N\rightarrow\infty$ we find
\bea
w^N_{i_N}/z_N&=&\tilde{w}^N_{i_N} \nn
w_i/z_N&\rightarrow& 0\nn
z_n/z_N&\rightarrow& 0
\eea
with $\tilde{w}^N_{i_N}$ finite. Also
\ben
B(w^N_{i_N})\sim \frac{x_N/u_N}{(1-\tilde{w}^N_{i_N})z_N}
\een
Hence
\bea
&&\lim_{z_N,x_N\rightarrow\infty}z^{\frac{2j_N(j_N+1)}{t}}x^{-2j_N}W_N^{(II)}
\nn
&\sim&z^{2\frac{j_N(j_N+1)}{t}}x_N^{-2j_N}x_N^{-\sum_{i_N}k^N_{i_N}}
u_N^{\sum_{i_N}k^N_{i_N}}z_N^{\sum_{i_N}k^N_{i_N}}\prod_{i_N}
(1-\tilde{w}^N_{i_N})^{k^N_{i_N}}\nn
&\cdot&\prod_{i_N < j_N}(\tilde{w}^N_{i_N}-\tilde{w}^N_{j_N})^
{2k^N_{i_N}k^N_{j_N}/t}z_N^{\sum_{i_N < j_N}2k^N_{i_N}k^N_{j_N}/t}\nn
&\cdot&\prod_{i_N,i}(\tilde{w}^N_{i_N})^{2k^N_{i_N}k_i/t}
z_N^{\sum_{i_N,i}2k^N_{i_N}k_i/t}\prod_{i_N}(1-\tilde{w}^N_{i_N})^
{2k^N_{i_N}j_N/t}z_N^{\sum_{i_N}2k^N_{i_N}j_N/t}\nn
&\cdot&\prod_{i_N,n}(\tilde{w}^N_{i_N})^{2k^N_{i_N}j_n/t}
z_N^{\sum_{i_N,n}2k^N_{i_N}j_n/t}z_N^{\sum_n2j_Nj_n/t}z_N^{\sum_i2j_Nk_i/t}\nn
&\cdot&u_N^{2j_N-1}e^{1/u_N}\Gamma(2j_N+1)
\dtp{u_N}W_N^{(I)}\prod_{i_N}d\tilde{w}^N_{i_N}
z_N^{\rho_N+\sigma_N}
\eea
We may evaluate the total power of $z_N$ as
\bea
&&\frac{2}{t}j_N(j_N+1)-2j_N+\frac{1}{t}{[}(\sum_{i_N}k^N_{i_N})^2-\sum_{i_N}
(k^N_{i_N})^2{]}+\frac{2}{t}(-2j_N)(-\sum_nj_n+j_N)\nn
&+&\frac{2}{t}j_N(-2j_N)+\frac{2}{t}(-2j_N)\sum_nj_n+\frac{2}{t}j_N\sum_nj_n
+\frac{2}{t}j_N(-\sum_nj_n+j_N)+\rho_N+\sigma_N\nn
&=&0
\eea
which merely shows that the intertwining field,
${[}\phi_{j_N}(z_N,x_N){]}^0_{j_N}$ indeed has the right scaling dimension
in our formalism. Similarly the total power of zero for $x_N$ says that we
treat the field in our formalism with correct global $sl_2$ properties.
The $u_N$-integrand becomes trivial, involving only
\ben
u_N^{-1}e^{1/u_N}
\een
The dependence on $\tilde{w}^N_{i_N}$ becomes
\bea
&&\prod_{i_N}(1-\tilde{w}^N_{i_N})^{k^N_{i_N}}\prod_{i_N<j_N}
(\tilde{w}^N_{i_N}-\tilde{w}^N_{j_N})^{2k^N_{i_N}k^N_{j_N}/t}\nn
&\cdot&\left (\prod_{i_N}(\tilde{w}^N_{i_N})^{2k^N_{i_N}/t}\right)
^{(-\sum_nj_n+j_N)}\prod_{i_N}(1-\tilde{w}^N_{i_N})^{2k^N_{i_N}j_N/t}\nn
&\cdot&\left (\prod_{i_N}(\tilde{w}^N_{i_N})^{2k^N_{i_N}/t}\right)^{\sum_nj_n}
\eea
So that this yields an integral over the $\tilde{w}^N_{i_N}$'s which is
independent of the remaining parameters of the correlator, except $j_N$, and
thus merely contributes to the normalization of the state $\bra{j_N}$.

\section{The 4-point function}
Using the general results for the $N$-point correlators, we now specialize
by way of illustration
to the case of 4-point correlation functions. First consider the $\beta\gamma$
part,
\ben
\bra{j_4}[\phi_{j_3}(z_3,x_3)]_j^{j_4}[\phi_{j_2}(z_2,x_2)]_{j_1}^j
\ket{j_1}_{\beta\gamma}
\een
with the notation
\bea
j_1+j_2-j&=&\rho_2-\sigma_2t\nn
j+j_3-j_4&=&\rho_3-\sigma_3t
\eea
So as far as the $\beta \g$ part is concerned this means we get
\bea
&&\oint_0\dtp{u_2}\dtp{u_3}\frac{1}{u_2u_3}u^{\rho_2-\sigma_2 t}\nn
&\cdot&\prod_{i_3=1}^{\rho_3}
{[}\frac{x_2/u}{w(3,i_3)-z_2}+\frac{x_3}{w(3,i_3)-z_3}{]}
\prod_{l_3=1}^{\sigma_3}
{[}\frac{x_2/u}{v(3,l_3)-z_2}
+\frac{x_3}{v(3,l_3)-z_3}{]}^{-t}\nn
&\cdot&\prod_{i_2=1}^{\rho_2}
{[}\frac{x_2/u}{w(2,i_2)-z_2}+\frac{x_3}{w(2,i_2)-z_3}{]}
\prod_{l_2=1}^{\sigma_2}
{[}\frac{x_2/u}{v(2,l_2)-z_2}
+\frac{x_3}{v(2,l_2)-z_3}{]}^{-t}\nn
&\cdot&(u_3^{-1}D_3)^{-2j_3+\rho_3-\sigma_3t}\exp\{\frac{1}{u_3}\}\G(2j_3+1)\nn
&\cdot&(u_2^{-1}D_2)^{-2j_2+\rho_2-\sigma_2t}\exp\{\frac{1}{u_2}\}\G(2j_2+1)\nn
\eea
where, we have let
\ben
u\equiv u_2/u_3
\een
Again the (somewhat misleading) notation is that the $D$'s are derivatives with
respect to the entire argument of the relevant exponentials.

Now we notice the following identity for fractional
derivatives of exponentials:
\ben
D_x^a\exp(x)D_y^b\exp(y)=D^{a+b}\exp(x+y)
\een
We want to consider a change of variables from $(u_2,u_3)$ to $(u_2,u)$.
The integration measure is
$$\frac{du_2}{u_2}\frac{du_3}{u_3}=-\frac{du_2}{u_2}\frac{du}{u}$$
\ben
2j_2+2j_3-\rho_3+\sigma_3t-\rho_2+\sigma_2t
=2j_2+2j_3-(j_3+j-j_4)-(j_2+j_1-j)=j_2+j_3+j_4-j_1
\equiv J_1
\een
We have introduced the notation
\ben
J_i\equiv j_1+j_2+j_3+j_4-2j_i
\een
Using the generalized exponential identity, we obtain the following $u$ and
$u_2$ dependence:
\ben
\frac{dudu_2}{uu_2}u^{-2j_3+\rho_2-\sigma_2t+\rho_3-\sigma_3t}
u_2^{J_1}D^{-J_1}\exp\{\frac{1+u}{u_2}\}
\een
Now the integral over $u_2$ will produce the factor
\ben
\frac{(1+u)^{J_1}}{\G(J_1+1)}
\een
We are left with the following integral:
\bea
&&\oint_0\dtp{u}
\prod_{i_3=1}^{\rho_3}{[}\frac{x_2}{w(3,i_3)-z_2}+\frac{x_3u}{w(3,i_3)-z_3}{]}
\prod_{l_3=1}^{\sigma_3}{[}\frac{x_2}{v(3,l_3)-z_2}+
\frac{x_3u}{v(3,l_3)-z_3}{]}^{-t}\nn
&\cdot&
\prod_{i_2=1}^{\rho_2}{[}\frac{x_2}{w(2,i_2)-z_2}+\frac{x_3u}{w(2,i_2)-z_3}{]}
\prod_{l_2=1}^{\sigma_2}{[}\frac{x_2}{v(2,l_2)-z_2}+
\frac{x_3u}{v(2,l_2)-z_3}{]}^{-t}\nn
&\cdot&u^{-2j_3-1}\frac{(1+u)^{J_1}}{\G(J_1+1)}\G(2j_3+1)\G(2j_2+1)
\eea

To write the final result for the 4-point function in a more compact form,
let us collectively denote the positions for both kinds of screening charges
as
$$ w_i, \ i=1,...,\rho_2+\rho_3+\sigma_2+\sigma_3\equiv M $$
Then the complete expression for the 4-point function is
\bea
&&\bra{j_4}[\phi_{j_3}(z_3,x_3)]_j^{j_4}[\phi_{j_2}(z_2,x_2)]_{j_1}^j
\ket{j_1}\nn
&=& \prod_{m<n}(z_m-z_n)^{2j_mj_n/t}\G(2j_3+1)\G(2j_2+1)\nn
&\cdot&\oint_0 \dtp{u}
\prod_{i=1}^{M}\oint \dtp{w_i}
{[}\frac{x_2}{w_i-z_2}+\frac{x_3u}{w_i-z_3}{]}^{-k_i}
\prod_{i<j}(w_i-w_j)^{\frac{2k_ik_j}{t}}\nn
&\cdot&
\prod_{i=1}^{M}\prod_{l=1}^{3}(w_i-z_l)^{\frac{2k_ij_l}{t}}
u^{-2j_3-1}\frac{(1+u)^{J_1}}{\G(J_1+1)}
\eea
where only one auxiliary integration over $u$ is involved.
Let us now work out some specific examples. In those example we shall fix
$z_3=x_3=1$. Due to projective invariance in both $z$ and $x$ spaces, the
more general case can be easily recovered.

Example 1:
\bea
j_1&=&j_2=j_3=j_4=1/2,\ \ j=0, \ \ J_1=1 \nn
\rho_2&=&1,\ \ \sigma_2=\rho_3=\sigma_3=0
\eea
In this case the $u$ integral can be explicitly carried out, leaving
the result
\bea
&&\bra{1/2}[\phi_{1/2}(1,1)]_0^{1/2}[\phi_{1/2}(z,x)]_{1/2}^0 \ket{1/2}\nn
&=&-((1-z)z)^{1/(2t)}\oint\dtp{w} (\frac{1}{1-w}+\frac{x}{z-w})
((1-w)(z-w)w)^{-1/t}\nn
&=& (-)^{1/t}((1-z)z)^{1/(2t)}
\left\{\binomial{-1/t}{1/t-1}z^{-2/t+1}F(1/t+1,-1/t+1;-2/t+2;z)\right.\nn
&+&\left.
x\binomial{-1/t-1}{1/t-1}z^{-2/t}F(1/t,-1/t+1;-2/t+1;z)\right\}
\eea
Here the $w$-integration contour goes through $z$ and encircles $w=0$, and
$F(a,b;c;z)$ is the hypergeometric function.

Example 2:
In this example we shall consider the case where a second screening charge
is needed.
\bea
j_1&=&-t/2,\ \ j_2=1/2,\ \ j_3=-t/2,\ \ j_4=1/2,\ \ j=1/2-t/2, \ \ J_1=1 \nn
\sigma_3&=&1,\ \ \sigma_2=\rho_2=\rho_3=0
\eea
In this case both the $u$ and $w$ integrals can be carried out, and yields
\bea
&&\bra{1/2}[\phi_{-t/2}(1,1)]_{1/2-t/2}^{1/2}[\phi_{1/2}(z,x)]_{-t/2}^{1/2-t/2}
\ket{-t/2}\nn &=&
(z(1-z))^{-1/2}\oint \dtp{w} \oint_0 \dtp{u}
(-\frac{u}{1-w}+\frac{x}{w-z})^{-t}\nn
&\cdot&(1+u)u^{t-1}(w-z)((w-1)w)^{-t}\G(-t+1)\nn
&=& (z(1-z))^{-1/2}(1-t+(t-2)z+xt)\frac{\G(1-t)(e^{-2\pi it}-1)}{(2-t)(1-t)}
\eea
where the $w$-integration contour may be taken to be the unit circle.

It can be verified that the expressions obtained in the two examples satisfy
the KZ equation.

We have also verified that these correlators respect the interesting conjecture
made in ref. \cite{FGPP} that in the limit $x\rightarrow z$ they reduce to
minimal model correlators. We hope to come back elsewhere and give a complete
discussion of the validity of this conjecture based on our full integral
representations.

\section{Proof of the Knizhnik-Zamolodchikov equations}

One may wonder whether the rules for contractions we have put forward, really
reproduce the structure of the conformal theory. In order to settle this
question in the affirmative we provide in this section an explicit proof that
our $N$-point functions satisfy the Knizhnik-Zamolodchikov equations. In this
proof we should not, therefore, make any use of the rules of contractions.

The Knizhnik-Zamolodchikov equation corresponding to the primary field at
position $z_{m_0}$ may be written down as:
\ben
\{t\pa_{z_{m_0}}+\sum_{m\neq m_0}
\frac{2D^a_{x_{m_0}}D^a_{x_m}}{z_m-z_{m_0}}\}W_N
=0
\label{KZ}
\een
where $W_N$ is the $N$-point function after requisite integrals have been
performed.

The structure of the proof is as follows:

For the selected position, $z_{m_0}$, of the primary field at that position, we
shall define a function
\ben
G(w)=\frac{1}{w-z_{m_0}}\{D^+_{x_{m_0}}G^-(w)+2D^3_{x_{m_0}}G^3(w)+
D^-_{x_{m_0}}G^+(w)\}
\een
where the $G^a(w)$'s are functions to be defined and will turn out a posteriori
to be
\ben
G^a(w)=\br J^a(w){\cal O}\kt
\een
where ${\cal O}$ is the collection of free field realizations of all our
chiral vertex operators and screening charges. Indeed we shall evaluate
the $G^a(w)$'s using our contraction rules from that idea. However the point
about the proof is that the function $G(w)$ eventually written down will only
have pole singularities as a function of $w$, and will behave as
${\cal O}(w^{-2})$
for $w\rightarrow\infty$, and thus the sum of residues will
vanish. What we shall show explicitly is that the vanishing condition for
this sum of residues is precisely the Knizhnik-Zamolodchikov equation on our
$N$-point function. This should come as no surprise since this is merely the
standard technique for proving the Knizhnik-Zamolodchikov equation. The point
is that in the standard proof one makes use of associativity properties of
the operators, and the purpose of our proof is exactly to establish that our
rules for contractions in fact conform to those.

Now it is clear how we build the functions $G^a(w)$. It is useful to observe
that to use our contraction rules with the free field realizations
\eq{wakimoto} is very easy in our correlator, since one may
establish the rules:
\bea
\beta(w)&\rightarrow&B(w)\nn
\gamma(w)&\rightarrow&-\sum_{i=1}^M\frac{D_{B_i}}{w-w_i}\nn
-\sqrt{t/2}\pa\varphi(w)&\rightarrow&\sum_{i=1}^M\frac{k_i}{w-w_i}+
\sum_{m=1}^{N-1}\frac{j_m}{w-z_m}
\eea
where
$$D_{B_i}\equiv\frac{\pa}{\pa B(w_i)},\ \ \ \ k_i \
\mbox{is defined as in section 5}$$
Then (our notation here does not distinguish between integrands and integrated
expressions)
\bea
G^+(w)&=&B(w)W_BW^\varphi_NF\nn
G^3(w)&=&\{B(w)\sum_{i=1}^M\frac{D_{B_i}}{w-w_i}+\sum_{i=1}^M\frac{k_i}{w-w_i}
+\sum_{m=1}^{N-1}\frac{j_m}{w-z_m}\}W_BW^\varphi_NF\nn
G^-(w)&=&\{-\sum_{i,j}B(w)\frac{D_{B_i}D_{B_j}}{(w-w_i)(w-w_j)}
+(t-2)\sum_i\frac{D_{B_i}}{(w-w_i)^2}\nn
&-&2\sum_{i,j}\frac{k_iD_{B_j}}{(w-w_i)(w-w_j)} -
2\sum_{m,j}\frac{j_mD_{B_j}}{(w-z_m)(w-w_j)}\}W_BW^\varphi_NF
\eea
These expressions define our function $G(w)$ and from now on we may completely
forget they came form applying our contraction rules to certain correlators.

In the following calculations the structure of the auxiliary $u$-integrations
turns out to be very crucial. In fact, any dependence on an $x_m$ is via
the combination $x_m/u_m$ so that we may write
$$x_m\pa_{x_m}=-u_m\pa_{u_m}$$
and subsequently do a partial integration in  $u_m$, writing effectively
$$x_m\pa_{x_m}W_BW^\varphi_NF\sim W_BW^\varphi_N\pa_{u_m}(u_mF)$$
Let
$$D^-_{x_{m_0}}G^+_{z_{m_0}}$$
denote the contribution to the pole residue
in $G(w)$ at $w=z_{m_0}$ coming from the term
$$D^-_{x_{m_0}}G^+(w)\equiv D^-_{x_{m_0}}\oint_{z_{m_0}}\dtp{w}
\frac{1}{w-z_{m_0}}G^+(w)$$
in $G(w)$ etc. Then we find after some calculations for the pole at
$z_{m_0}$:
\bea
D^-_{x_{m_0}}G^+_{z_{m_0}}&=&\sum_{\ell\neq m_0}
\frac{x_\ell/u_\ell}{z_{m_0}-z_\ell}\pa_{x_{m_0}}W_B\cdot W^\varphi_NF\nn
2D^3_{x_{m_0}}G^3_{z_{m_0}}&=&2\{-\sum_{\ell\neq m_0}\frac{x_\ell/u_\ell}
{z_{m_0}-z_\ell}u_{m_0}\pa_{x_{m_0}}W_B\cdot W^\varphi_N\nn
&-&\pa_{z_{m_0}}W_B\cdot W^\varphi_N
+\frac{t}{2j_{m_0}}W_B\cdot\pa_{z_{m_0}}W^\varphi_N\}
(-j_{m_0}+u^{-1}_{m_0})F\nn
D^+_{x_{m_0}}G^-_{z_{m_0}}&=&\{-\sum_{\ell\neq m_0}\frac{x_\ell/u_\ell}
{z_{m_0}-z_\ell}\pa_{x_{m_0}}W_B\cdot W^\varphi_N{[}2j_{m_0}u_{m_0}^{2j_{m_0}}
-u_{m_0}^{2j_{m_0}-1}{]}\nn
&-&2\pa_{z_{m_0}}W_B\cdot W^\varphi_N{[}(2j_{m_0}-1)u_{m_0}^{2j_{m_0}-1}-
u_{m_0}^{2j_{m_0}-2}{]}\nn
&+&(t-2+2j_{m_0})\pa_{z_{m_0}}W_B\cdot W^\varphi_N u_{m_0}^{2j_{m_0}-1}\nn
&+&\frac{t}{j_{m_0}}W_B\cdot \pa_{z_{m_0}}W_N^{\varphi}
{[}2j_{m_0}u_{m_0}^{2j_{m_0}-1}-u_{m_0}^{2j_{m_0}-2}{]}\}\nn
&\cdot&e^{1/u_{m_0}}\Gamma(2j_{m_0}+1)\prod_{m\neq m_0}\Gamma(2j_m+1)
u_m^{2j_m-1}e^{1/u_m}
\eea
This sums up to become
$$t\pa_{z_{m_0}}W_N$$
the first term in the Knizhnik-Zamolodchikov equation.

In a similar fashion the pole residue at $w=w_j$ turns out to give after
some calculations (up to a factor independent of $w_j$)
$$\pa_{w_j}
\left (\frac{tW_B\cdot W^\varphi_N}{B(w_j)(w_j-z_{m_0})}\right )$$
so that this term will vanish upon integration over $w_j$.

Finally the pole residues at the points $w=z_m\neq z_{m_0}$ rather
easily give the remaining terms in the KZ equations, \eq{KZ}. Since there
are no other singularities in $w$, the KZ equations have been proven.

\section{Projective invariance and global $SL(2)$ invariance}
In ref. \cite{DF} it was shown, that solutions of the Knizhnik-Zamolodchikov
equations are projectively invariant provided the primary fields can add up
to a singlet. In our case this is the requirement of global $SL(2)$ invariance.
Thus, let us restrict to the case where the initial and final bra and ket
carry just the vacuum and the dual vacuum: $j_1=0=j_N$, i.e. we are really
looking at an $(N-2)$-point function.
Then global $SL(2)$ invariance is the statement that
\ben
\sum_{m=2}^{N-1}D^a_{x_m}W_{N-2}=0
\een
This is equivalent to the statement that in
\ben
G^a(w)=\br J^a(w){\cal O}\kt
\een
the leading behaviour as $w\rightarrow\infty$ is ${\cal O}(w^{-2})$ rather than
${\cal O}(w^{-1})$. From the expressions above, that is trivial for $G^-(w)$,
and for $G^3(w)$ it follows from the fact that
\ben
\sum_{i=1}^Mk_i=-\sum_{m=1}^{N-1}(\rho_m-t\sigma_m)=-\sum_{m=1}^{N-1}j_m
\een
for $j_N=0$.
For $G^+(w)$ it is more complicated to see. As previously discussed this is
related to the fact that we are not using the projectively invariant vacuum,
$\bra{sl_2}$, in our calculations, rather we are using the dual vacuum,
$\bra{0}$, for which $\bra{0}J^+_0=\bra{0}\beta_0\neq 0$.

What we are going to show is that the state,
$$\bra{0}\beta_0$$
even though it is non-vanishing, is BRST-exact in the sense of Felder \cite{F},
and that (hence) it decouples from
correlators of BRST invariant operators.

First let us argue at the operator level, and subsequently at the level of our
correlators. We write
\ben
\bra{0}\beta_0=\bra{0}e^{-\sqrt{2/t}q_\varphi}e^{\sqrt{2/t}q_\varphi}\beta_0
=\bra{-1}\oint\dtp{z}e^{\sqrt{2/t}\varphi(z)}\beta(z)
\een
where the bra state, $\bra{-1}$, is the lowest weight state, $\bra{j=-1}$.
Now this integral is in fact the appropriate BRST operator in
Felder's formulation.
To see this, recall, that acting on the Fock space pertaining to
$$j_{r,s}$$
and labelled $F_{r,s}$, the relevant BRST operator (for which the BRST current
is single valued) is
\ben
Q_r\sim \oint\dtp{v_0}...\dtp{v_{r-1}}S_1(v_0)...S_1(v_{r-1})
\een
with
$$Q_r: \ F_{r,s}\mapsto F_{-r,s}$$
For $j_N$ =0, this is the Fock space with $r=1,s=0$, so that in fact
$$2j_N+1=1-0\cdot t$$
therefore the relevant BRST operator on this space is $Q_1$ which is just the
one we obtained.

Next let us see how the argument works at the level of the correlator. We are
going to show that inserting the operator
$$\beta_0=\oint_\infty\dtp{w}\beta(w)$$
furthest to the left in a correlator with $j_N=0$, is equivalent to inserting
the BRST-charge operator
$$\oint_\infty\dtp{w}\beta(w)e^{\sqrt{2/t}\varphi(w)}$$
furthest to the left of the operators and to the right of the bra $\bra{-1}$.

Indeed in the first case, using the rules for building correlators, we obtain
(up to normalization)
\bea
&&\oint_\infty
\dtp{w}B(w)\prod_iB(w_i)^{-k_i}\prod_{i<j}(w_i-w_j)^{2k_ik_j/t} \nn
&\cdot&\prod_{n<m}(z_n-z_m)^{2j_nj_m/t}\prod_{i,n}(w_i-z_n)^{2k_ij_n/t}
\prod_n u_n^{2j_n-1}e^{1/u_n}
\eea
In the second case, we notice that the state labelled $\bra{-1}$ is exactly
a lowest weight state, $\bra{j_N=-1}$, and we are formally looking
at an $(N-1)$ point function with $j_N=-1$. Since we have
\ben
\iota_{N-2}+j_{N-1}-j_N=\rho_{N-1}-\sigma_{N-1}t
\een
we see that the value of $\rho_{N-1}$ will be one unit larger in the case
$j_N=-1$ than in the case $j_N=0$. This means we have one extra screening
charge of the first kind in that case, which we may lift off the intertwining
field furthest to the left, in other words that calculation will just be the
one we seek to carry out.
Using the rules developed we find in that case, letting $w$ denoting the
position of the extra screening operator compared to the previous case (so in
the formulas below, the index, $i$, runs over exactly the same set as before):
\bea
&&\oint_\infty\dtp{w}B(w)\prod_i B(w_i)^{-k_i}\prod_i (w-w_i)^{-2k_i/t}
\prod_n(w-z_n)^{-2j_n/t}\prod_{i<j}(w_i-w_j)^{2k_ik_j/t}\nn
&\cdot&\prod_{n<m}(z_n-z_m)^{2j_nj_m/t}\prod_{i,n}(w_i-z_n)^{2k_ij_n/t}
\prod_nu_n^{2j_n-1}e^{1/u_n}\nn
&=&\oint_\infty\dtp{w}B(w)\prod_i B(w_i)^{-k_i}\prod_i (1-w_i/w)^{-2k_i/t}
\prod_n(1-z_n/w)^{-2j_n/t}  w^{-2(\sum_ik_i+\sum_nj_n)/t}
\nn
&\cdot&\prod_{i<j}(w_i-w_j)^{2k_ik_j/t}\prod_{n<m}(z_n-z_m)^{2j_nj_m/t}
\prod_{i,n}(w_i-z_n)^{2k_ij_n/t}\prod_n u_n^{2j_n-1}e^{1/u_n}
\eea
We now use that
\ben
-\sum_ik_i =\sum_n\rho_n-t\sum_n\sigma_n=\sum_nj_n-j_N
\een
Since we use the notation that sums over $i$ and $n$ are pertaining to the
case with $j_N=0$, we see that the extra power of $w$ becomes zero, and for
a very large integration contour for $w$, all $w$-dependence drops out except
for the one in $B(w)$ so that we exactly prove the identity of the two cases
also at the level of our correlators.

Having come this far, we may move the Felder-type BRST operator through
all intertwining fields in exactly the same way as for minimal models, until
in the end it hits the vacuum. Since we are only using screening operators of
the first kind in the BRST operators, the procedure will work just as for the
minimal models.

In addition it is rather easy to verify that the functions $G^a(w)$ have
the expected pole residues for $\br J^a(w){\cal O}\kt$ at points
$w=z_m$. Further one verifies that there are no pole residues in $G^+(w)$
and $G^3(w)$ for $w=w_i$ the position of a screening charge. For $G^-(w)$
that residue is proportional to the total derivative
$$\pa_{w_i}\left (D_{B_i}W_B\cdot W^\varphi_N\right )$$
All of those remarks establish that our $N$-point blocks have the correct
projective and global $SL(2)$ invariance properties. In particular, it gives
rise to alternative representations for the $N$-point blocks, ones obtained
by taking initial and final states to be vacua. This new representation is
more symmetric and does not involve dual states \cite{D90}, but involves
a larger number of screening integrations in general.

\section{Conclusions}
In this paper we have shown how to deal with fractional powers of free fields,
and thereby we have managed to make sense of the second screening charge
proposed by Bershadsky and Ooguri \cite{BO} for affine $SL(2)$ WZNW models.
This has enabled us to build the most general conformal blocks for such
theories on the sphere, even in the case of admissible representations with
fractional levels. The ensuing integral formulas are tractable and have
allowed us to verify many formal
properties, such as projective invariance, global $SL(2)$ invariance, the fact
that the conformal blocks satisfy the Knizhnik-Zamolodchikov equations etc.
This gives hope that the relations to non perturbative string theory and
topological field theory along the
lines of \cite{HY,AGSY} may be made more explicit. Also our technique
seems straightforward to generalize to higher groups and supergroups
so that further progress in non
perturbative string theory conceivably could be obtained.

\appendix
\section{Fractional Calculus}
Here we very briefly introduce fractional calculus \cite{MR}.
For an analytic function, $f(z)$, the fractional derivative, $\pa_z^a$,
is defined for any complex number, $a$. It satisfies the following axioms:
\begin{enumerate}
\item
If $f(z)$ is an analytic function, the fractional derivative,
$\pa^{\delta}f(z)$, is an analytic function of $z$ and $\delta$.
\item For $\delta$ integer, the result must agree with ordinary differentiation
($\delta$ positive) or integration ($\delta$ negative). By default the
integration constants are put to zero, so that the function together with
a maximum number of derivatives vanish at some point, like $z=0$.
\item $\delta=0$ is the identity.
\item Fractional differentiation is linear.
\item For fractional integration, $\Re\alpha>0,\Re\beta>0$
$$\pa^{-\alpha}\pa^{-\beta}f(z)=\pa^{-(\alpha+\beta)}f(z)$$
\end{enumerate}

For $\Re a>0$, $\pa^{-a}$ satisfying the above is  given by the
Riemann-Liouville operator
\ben
\pa^{-a}f(z)=\frac{1}{\Gamma(a)}\int_0^z(z-t)^{a-1}f(t)dt
\een
For any $a,b$
\ben
\pa^a\pa^b=\pa^{a+b}
\een
Thus fractional differentiation is also defined.
We principally need the rule, which is now easy to derive
\ben
\pa^a_x x^b=\frac{\Gamma(b+1)}{\Gamma(b-a+1)}x^{b-a}
\een
(Notice that $\pa_x^a 1\neq 0$ for $a$ not positive integer (!)).

\section{Example of consistency conditions for Wick contractions}
In this appendix we illustrate the non-trivial nature of the workings of the
Wick contractions we have proposed. In fact, consider the reduction of
\ben
\beta^a(z)\gamma^a(z')=\frac{\Gamma(a+1)}{(z-z')^a}\sum_N\binomial{a}{N}
\frac{1}{N!}(z-z')^N:(\beta(z)\gamma(z'))^N:
\label{baga}
\een
where we have used the rules developed in sects. 2 and 3. Next consider
the evaluation of
\ben
I_{a,b}(z,z')\equiv (\beta^a(z)\gamma^a(z'))(\beta^b(z)\gamma^b(z'))
\een
This expression may be evaluated in two ways: either (i) by first using
\eq{baga}
for both parentheses to reduce both of them to normal ordered products of
integer powers of $\beta$ and $\gamma$, and then subsequently carrying out all
remaining contractions, or else (ii) by simply using \eq{baga} with $a+b$
replacing $a$. Obviously these two ways should lead to the same result
for consistency. This requirement is part of the associativity properties
for operators, and we consider them to be justified by the fact that our
blocks satisfy the KZ equations. Here we demonstrate that the above condition
gives rise to nontrivial identities for which we indicate an independent
elementary proof.

The general contractions between integer powers of $\beta$ and $\gamma$ are
carried out using the following trick:
\bea
\beta^n(z)\gamma^m(z')&=&{[}\beta(z)+(z-z')^{-1}\pa_{\gamma(z')}{]}^n
\gamma^m(z')\nn
&=&:\exp\{(z-z')^{-1}\pa_{\gamma(z')}\pa_{\beta(z)}\}\beta^n(z)\gamma^m(z'):
\nn
&=&\sum_\ell\frac{(z-z')^{-\ell}}{\ell!}:\pa^\ell_{\beta(z)}\beta^n(z)
\pa^\ell_{\gamma(z')}
\gamma^m(z'):\nn
&=&\sum_\ell \frac{1}{(z-z')^{\ell}\ell!}\frac{\Gamma(n+1)}{\Gamma(n-\ell +1)}
\frac{\Gamma(m+1)}{\Gamma(m-\ell +1)}:\beta^{n-\ell}(z)\gamma^{m-\ell}(z'):
\eea
Then one obtains after a few steps
\bea
I_{a,b}(z,z')&=&\frac{\Gamma(a+1)\Gamma(b+1)}{(z-z')^{a+b}}
\sum_{m,n,k,\ell}\binomial{a}{m}\binomial{b}{n}\binomial{m}{k}
\binomial{n}{\ell}\nn
&\cdot&\frac{:(\beta(z)\gamma(z')(z-z'))^{m+n-k-\ell}:}{(m-\ell)!(n-k)!}
\eea
On the other hand the second way of evaluation simply gives the result
\ben
I_{a,b}(z,z')=\frac{\Gamma(a+b+1)}{(z-z')^{a+b}}\sum_N\binomial{a+b}{N}
(z-z')^N\frac{:(\beta(z)\gamma(z'))^N:}{N!}
\een
Defining the generating functions
\bea
F_a(x)&=&\sum_N\binomial{a}{N}\frac{x^N}{N!}\nn
G_{ab}(x)&=&\sum_{m,n,k,\ell}\binomial{a}{m}\binomial{b}{n}\binomial{m}{k}
\binomial{n}{\ell}\frac{x^{m+n-k-\ell}}{(m-\ell)!(n-k)!}
\eea
we see that the consistency condition may be expressed as
\ben
G_{ab}(x)=\binomial{a+b}{b}F_{a+b}(x)
\label{cc}
\een
We now briefly indicate how this identity may be proven. First we notice
that we may write
\ben
\binomial{a+b}{a}=\binomial{a+b}{b}=\oint\dtp{t}\frac{(1+t)^{a+b}}{t^{a+1}}
\een
where the contour may be taken as the unit circle, passing through the
branch point at $t=-1$ of the integrand (for suitable values of the exponents).
Then we write
\bea
G_{ab}(x)&=&\sum_{p,q,k,\ell}\binomial{a}{q+\ell}\binomial{b}{p+k}
\binomial{q+\ell}{k}\binomial{p+k}{\ell}\frac{x^{p+q}}{p!q!}\nn
&=&\sum_{p,q,k,\ell}\prod_{i=1}^4\oint_{{\cal C}_0}\dtp{t_i}\frac{1}{t_i}
\frac{(1+t_1)^a}{t_1^{q+\ell}}\frac{(1+t_2)^b}{t_2^{p+k}}
\frac{(1+t_3)^{q+\ell}}{t_3^k}\frac{(1+t_4)^{p+k}}{t_4^\ell}\frac{x^{p+q}}{p!q!}
\eea
The identity is now obtained by successively doing 1) the sum over $\ell$,
2) the integral over $t_4$, 3) the sum over $k$, 4) the integral over $t_3$,
5) the integral over $t_1$ and $t_2$ in any order, and 6) the sum over $q$
with $p+q=N$.
In each case one picks up
residues after suitable deformations of the contours.\\[.5cm]
{\bf Acknowledgement} \\[.2cm]
We are indebted to Anna Tollst\'en for discussions at the
early stages of this work.
Ming Yu would like to thank J.-B. Zuber and F.G. Malikov for private
communications. M.Y. also thanks the Danish Research Academy for financial
support.

\end{document}